\documentclass[sigconf]{acmart}

\usepackage{graphicx}
\usepackage{multirow}
\usepackage{subcaption}
\usepackage{xcolor}
\usepackage[draft]{fixme}

\setcopyright{acmlicensed}
\copyrightyear{2026}
\acmYear{2026}
\setcopyright{cc}
\setcctype{by}
\acmConference[CSLAW '26]{Symposium on Computer Science and Law}{March 03--05, 2026}{Berkeley, CA, USA}
\acmBooktitle{Symposium on Computer Science and Law (CSLAW '26), March 03--05, 2026, Berkeley, CA, USA}
\acmPrice{}
\acmDOI{10.1145/3788646.3789523}
\acmISBN{979-8-4007-2447-3/2026/03}

\fxsetup{layout=inline,theme=color,mode=multiuser,inlineface=\itshape,envface=\itshape}
\FXRegisterAuthor{sv}{asv}{\colorbox{gray!10!white}{\color{black}Suresh}}
\FXRegisterAuthor{jw}{ajw}{\colorbox{blue!10!white}{\color{black}Jenn}}
\FXRegisterAuthor{as}{aas}{\colorbox{green!10!pink}{\color{black}Andrew}}
\FXRegisterAuthor{sb}{asb}{\colorbox{green!10!pink}{\color{black}Solon}}

\begin{document}

\title{Distinguishing Task-Specific and General-Purpose AI in Regulation}

\author{Jennifer Wang}
\affiliation{
  \institution{Stanford University}
  \city{Stanford}
  \state{CA}
  \country{USA}
}
\email{jennjwang@stanford.edu}

\author{Andrew D. Selbst}
\affiliation{
  \institution{University of California, Los Angeles}
  \city{Los Angeles}
  \state{CA}
  \country{USA}
}
\email{aselbst@law.ucla.edu}

\author{Solon Barocas}
\affiliation{
  \institution{Microsoft Research}
  \city{New York City}
  \state{NY}
  \country{USA}
}
\email{solon@microsoft.com}

\author{Suresh Venkatasubramanian}
\affiliation{
  \institution{Brown University}
  \city{Providence}
  \state{RI}
  \country{USA}
}
\email{suresh@brown.edu}

\begin{abstract}
Over the past decade, policymakers have developed a set of regulatory tools to ensure AI development aligns with key societal goals. Many of these tools were initially developed in response to concerns with task-specific AI and therefore encode certain assumptions about the nature of AI systems and the utility of certain regulatory approaches. With the advent of general-purpose AI (GPAI), however, some of these assumptions no longer hold, even as policymakers attempt to maintain a single regulatory target that covers both types of AI.

In this paper, we identify four distinct aspects of GPAI that call for meaningfully different policy responses. These are the generality and adaptability of GPAI that make it a poor regulatory target, the difficulty of designing effective evaluations, new legal concerns that change the ecosystem of stakeholders and sources of expertise, and the distributed structure of the GPAI value chain.

In light of these distinctions, policymakers will need to evaluate where the past decade of policy work remains relevant and where new policies, designed to address the unique risks posed by GPAI, are necessary. We outline three recommendations for policymakers to more effectively identify regulatory targets and leverage constraints across the broader ecosystem to govern GPAI.
\end{abstract}



\maketitle
\renewcommand\footnotetextcopyrightpermission[1]{}


\section{Introduction}

Recent advances in AI have drawn intense interest from policymakers. In the past year alone, policymakers have proposed and enacted a litany of governance measures, from the European Union AI Act to a growing number of state and local initiatives in the United States \cite{EUAIAct2024, curry2024state}. This wave of legislative activity has generated new confusion and contestation over how AI should be defined and governed. Some treat generative and general-purpose AI (GPAI) as a fundamentally new class of technology whose scale and autonomy distinguish it from others \cite{bostrom2014superintelligence, altman2023governance}, advocating for entirely new laws and institutions \cite{nichols2023unaiwatchdog, bis2025ai_framework}. Others caution against reinventing the regulatory wheel, instead favoring the adaptation of existing laws \cite{corrigan2024governance}.\looseness=-1

A notable source of this confusion is the failure to distinguish between task-specific systems and GPAI. Some regulatory efforts draw this line explicitly -- for instance, the E.U. AI Act imposes distinct obligations on GPAI providers \cite{BoineRolnick2023WeRobot}. Many state and local proposals, however, do not. New York’s A6453-B, for example, defines AI as a system that ``can … make predictions, recommendations, or decisions,'' yet its provisions targeting model-weight theft are clearly aimed at generative systems \cite{NYAssembly2026A6453B}. We also see the reverse: laws designed around task-specific systems being imposed on GPAI. New Mexico HB 60 describes AI as a technology that can be used to ``generate outputs, including content, decisions, predictions, or recommendations,'' yet it requires the disclosure of ``sources or owners of the datasets'' and the ``factors in the data'' used to produce outputs --- assumptions that fit task-specific models with identifiable datasets and features, but not GPAI models \cite{ChandlerEtAl2025NM}.

This confusion matters because, over the past decade, policymakers have designed their regulatory toolkit around particular features of task-specific AI, but many of these features lose their relevance in a generative setting. For years before ChatGPT was released to the public, regulators have applied sector-specific frameworks to address the use of task-specific systems in employment, credit, housing, and public benefits \cite{cfpb, hud2024screening, jillson2021aiming}. 
These regimes were built around assumptions of how these systems are trained and evaluated for well-defined tasks, and how they are deployed to specific domains with legal guardrails. GPAI, however, breaks many of these assumptions. \looseness=-1

In this paper, we identify four important differences between general-purpose and task-specific AI that policymakers should consider in governing both types of systems. First, the generality and adaptability of GPAI makes it a poor regulatory target: a single model can be repurposed across many domains and use cases, so rules aimed at the model itself may fail to track where risk actually arises. Second, GPAI systems are difficult to evaluate in reliable ways because they lack task specificity. Third, GPAI raises new legal concerns that were absent or far less salient in task-specific settings, shifting the ecosystem of stakeholders and expertise required for oversight. Fourth, GPAI is developed and deployed through a highly distributed value chain which complicates the attribution of responsibility. These distinctions create governance challenges that necessitate new regulatory tools to enforce both existing and new laws.\looseness=-1

In light of these distinctions, policymakers will need to evaluate where the past decade of policy work remains fit for purpose and where new policies are needed to address the risks unique to generative AI. Consequently, we offer three recommendations. First, that regulation should target use cases and impact, rather than the technical specification of AI systems. Second, that technical proxies should be used sparingly in regulation, with a careful balancing of risk, benefits, and a consideration of alternatives. Third, that regulation should focus on the entire \emph{risk chain} for a use case and reinforce structural constraints that can limit harms, rather than focusing on AI systems alone.

\section{Conceptual Distinctions}
\label{sec:technical_distinctions}


In this paper, we draw a distinction between models built to perform specific tasks, typically referred to as \emph{task-specific}, \emph{predictive}, or \emph{discriminative}, and those constructed without a specific task in mind, often referred to as \emph{general-purpose} (including those that are later fine-tuned for a specific task) or \emph{generative}. While the distinction we raise may not capture every technical detail, it reflects the features most pertinent to policymakers.\footnote{The EU AI Act, among others, makes this same distinction, but calls out the former systems as `AI' and the latter as `general-purpose AI'. \citet{bommasani2021opportunities} refer to the former as `task-specific' (and the latter as `task-agnostic', but settle on `foundation'). We feel that the terms `task-specific' and `general-purpose' are useful for the distinctions that we wish to explore, but note that there are several different ways to refer to these systems, and there is no clear consensus.} 


 
\subsection{Task-specific AI systems}
Task-specific AI systems are designed to \emph{predict} a particular outcome and are almost exclusively built using supervised learning. 
Building such a system starts with a \emph{task} described by training data consisting of input vectors and the ``ground truth'' predictions. The goal is to build a system that can correctly reproduce the ground truth predictions given the inputs, in the hope that it will accurately predict outcomes for a new input. 

While this description is an oversimplification of model training, 
it captures the most salient aspects of task-specific AI for policymakers---there is an intended task, the task can be expressed in terms of specific training data, and the goal is to build a system that performs this task. 
\cite{selbst2019fairness}.


\subsection{General-purpose AI systems}

When we turn to GPAI, the picture is much more complicated. Some policy definitions
draw distinctions from task-specific systems based on the nature of the output---content versus predictions \cite{oecd2024genai}. This is reflected in AI definitions in legislative proposals that insert the word `content' into the list of outputs that might be produced by an AI system \cite{rhodeisland2025s0627, texas2025hb149, iowa2025hf406}.

Others focus on the development process and nature of the model. For example, \citet{bommasani2021opportunities} describe it as a model built using \emph{self-supervised learning} and that has billions of parameters. \citet{lee2024talkinboutaigeneration} rejects the term ``foundation model'' in favor of ``generative AI'', which they describe as characterized by a mixture of models, built at scale, that makes use of specific components like transformers.

The EU AI Act defines ``general purpose AI'' as ``an AI model, including where such an AI model is trained with a large amount of data using self-supervision at scale, that displays significant generality and is capable of competently performing a wide range of distinct tasks regardless of the way the model is placed on the market and that can be integrated into a variety of downstream systems or applications'' \cite{EUAIAct2024}. This definition retains the method of production (self-supervision), the reference to scale, and the implied capabilities.\looseness=-1

In line with these policy discussions, we view a GPAI system as one that is built using self-supervision, that usually (though not necessarily) is described using billions of parameters, and that most importantly is not trained to perform any specific task beyond that which the self-supervision requires.\footnote{To illustrate the difficulty even with this rendition, systems like BERT \cite{devlin2019bert} are built using self-supervision, but are not generative in the sense of producing new, fluent sequences of text.}
Self-supervision is perhaps the most distinctive feature of what characterizes GPAI models. In language models, for example, self-supervision proceeds by training the language model to fill in the blanks in lines of text in which words have been masked randomly.\footnote{The model is trained using supervised learning with a specific task---predict the missing words---but using the training data itself as inputs and targets for training. This gives rise to the term ``self-supervision.''
For image models, the process is different, but is still self-supervised.} The rationale for this form of training is that for a model to be able to correctly predict missing words in arbitrary sentences in text, it must be able to build an internal representation of meaning in text that will allow it to complete the sentences. Regardless of whether models are able to represent `meaning', it is the case that models do end up constructing an internal representation space for words (or tokens) and use this internal representation to respond to prompts or other natural language queries. This internal representation is what allows GPAI systems to perform a wide range of downstream tasks for which it might not have ever been trained specifically.\footnote{While some self-supervised systems are built for specific tasks, policy discussions focus on GPAI systems where self-supervision enables broad, task-agnostic capabilities.}

Most of the other elements used to define GPAI systems---the architecture of transformers, the apparent need for billions of parameters, and so on---are either consequences of the training process or reflections of prevailing design practices. These features represent our current understanding of how to build models effectively, but they are not necessary to capture the task-agnostic nature of GPAI systems.

But what of the myriad further steps in training a GPAI model, like fine-tuning, reinforcement learning from human feedback, inference-tuning, and so on? First, most of these post-pretraining steps do not fundamentally change the fact that a GPAI model relies on an internal representation to perform a wide range of downstream tasks, even if they might customize the model to perform more effectively on certain tasks. Second, these downstream training steps apply to an already built model, rather than creating a new one from scratch. Unlike task-specific models, where using the model for a new task usually involves rebuilding it from scratch, fine-tuning merely modifies an existing (and functioning) model, and can be performed without knowledge of how the original model was built.\footnote{The knowledgeable reader will note that transfer learning is exactly the process of taking a given model and adapting it using small amounts of new data for a related task. While this is true, it is typically less common in predictive settings to employ transfer learning since it is more straightforward to rebuild the model.} In turn, the builders of the base model need not have any knowledge of how fine-tuning might occur and for what purpose. A GPAI model that is fine-tuned to perform a prediction task is still a GPAI model for the purpose of the points we discuss in this paper. \looseness=-1
\section{What makes general-purpose AI distinctive?}

In this section, we outline four key distinctions that set GPAI apart---its general-purpose design, challenges in evaluation, emergence of new legal concerns, and distributed value chain structure. These differences challenge the assumptions underlying existing regulatory instruments, revealing how oversight and accountability mechanisms designed for task-specific systems lose their utility when applied to GPAI systems.

\subsection{General-purpose AI systems make poor regulatory targets because of their generality and adaptability.}
\label{sec:generality}

GPAI systems are general-purpose and adaptable, able to handle unstructured inputs across many modalities (e.g., text, images, audio, etc.).
As a result, the space of potential applications for GPAI is vast.
Through fine-tuning and prompting, the same base model can be adapted to perform remarkably different tasks without specific training. For example, GPT-4
can understand and generate human-like text across various topics and styles, process and generate images, engage in real-time verbal conversations, perform complex reasoning tasks, and write and debug code, all without fundamentally changing the underlying model \cite{openai2023}.

In contrast, task-specific systems are purpose-built to perform a defined task. They streamline, support, and sometimes fully automate human decision-making processes. Although these systems vary in their technical implementation, they share a common function as tools for decision-making. This shared function gives rise to a shared set of risks across different applications. Whether applied to hiring, credit scoring, or tenant screening, task-specific AI raises a similar set of concerns related to fairness, transparency, and accountability.
This creates a coherent basis for regulators to establish overarching guardrails that target universal harms posed by task-specific AI---one that draws on principles of due process, non-discrimination, and privacy. For example, the Blueprint for an AI Bill of Rights outlines common expectations for automated decision systems, such as the right to notice and explanation and protection from algorithmic bias, regardless of the domain \cite{wh_ostp_2022_ai_bill_of_rights}. Similarly, the proposed AI Civil Rights Act seeks to prohibit the use, sale, or promotion of algorithmic decision-making systems that discriminate on the basis of protected characteristics \cite{markey2024aicivilrights}. The comprehensive AI regulation passed in Colorado in 2024 similarly sets requirements for developers of "high-risk" AI systems across a wide range of domains \cite{coloradoSB205}.

The generality and adaptability of GPAI breaks down this regulatory coherence. 
GPAI systems can be put to a far more diverse set of uses and 
although these applications may stem from the same technical foundation, many will raise distinct risk profiles. Like task-specific AI, GPAI can be used for decision-making, including in the high-stakes context of hiring \cite{Gaebler_Goel_Huq_Tambe_2025}, lending \cite{BowenIII_Price_Stein_Yang_2024}, and housing \cite{Liu_So_Hosoi_DIgnazio_2024}, thus invoking the same set of regulatory concerns. But it can also be used to generate children’s stories, which raises concerns over age-appropriate material and creative rights \cite{unicef2021children, uscopyright2024ai}, for drafting legal documents, where there may be concerns over hallucinations and the confidentiality risks of client information \cite{ca_state_bar}, and for coding assistance, which could introduce security vulnerabilities and lower code quality \cite{Perry_2023, harding2025aicodequality}. While this generality and adaptabiltiy is not inherently problematic, as there are sector-specific frameworks that govern each of these use cases, the challenge arises when treating GPAI as the regulatory target (i.e., the specific entity that regulation is designed to govern).

Because there is no shared regulatory objective (i.e., the specific goal or intended outcome that regulation aims to achieve) across generative use cases, efforts to develop regulations at the model level may be misguided.
Model-level regulations that apply across diverse applications would require abstracting away all of the details that actually account for the concerns raised by these specific applications. Because GPAI systems lack a shared purpose, the only features shared across applications tend to be technical properties of the models, such as compute usage and model size, which do not correspond to any particular applications of concern. As a result, when regulators attempt to pursue broad upstream controls specifically targeting GPAI, they are left with few footholds other than these technical properties. These features are attractive from a regulatory perspective because they are easy to define and measure. However, they are often poor proxies for the many harms that could arise from different applications. 

Thus, regulation targeting GPAI systems often yields rules that are broad in scope, but shallow in substance. These rules are \textit{shallow} because they only address the highest-level factors that could affect the risks posed by GPAI---typically only those observable at the model level without consideration of the deployment context. This creates a regulatory framework that appears expansive but lacks the specificity to ensure meaningful oversight. We elaborate on the limitations of regulating at the model level in \ref{rec1}, where we instead recommend regulating the applications and impacts of these systems.
\looseness=-1

\subsection{General-purpose AI systems are difficult to evaluate due to their lack of task specificity.}
\label{eval}

Evaluation of AI systems has become a touchstone for oversight.  
In many cases, evaluations are designed to assess system performance using quantitative metrics, such as accuracy and recall, that typically require a known ground truth for comparison. By systematically assessing AI system performance, evaluations provide enforcement mechanisms and remedies to contest false claims and ineffective systems \cite{Raji2022Fallacy,NTIA2024}. They are crucial to determining whether a system is fit for purpose and sufficiently reliable for deployment. In other cases, evaluations are designed to assess models' impacts. One common example is the use of evaluation to identify when task-specific AI systems produce disparate impacts---that is, when they differentially benefit or burden different groups. From the E.U. AI Act to New York City local laws, policymakers have called for independent audits, impact assessments, and risk assessments in order to understand both basic functionality (i.e., performance) and harms (i.e., impacts), including discrimination, privacy, and physical safety \cite{EUAIAct2024, NYCInt1894_2020, WhiteHouseAIBoR_Safe2022}.
The widespread adoption of audit requirements and impact assessments in U.S. congressional bills \cite{AIA_2023}, state-level initiatives \cite{CaliforniaSB36, ColoradoSB169}, and international proposals \cite{EUAIAct2024, CanadaAIDA2023, ICO2023Fairness} reflects their importance in providing assurance and accountability in AI systems.

Task-specific systems are designed around tasks that map from specific input spaces to output spaces (see Section~\ref{sec:technical_distinctions}). This means developers are able to conduct evaluations within bounded problem domains where decision-making objectives can be clearly defined and measured. Typically, evaluators use a holdout set---data withheld from training that contains known outcomes that serve as the ground truth for assessing the quality of the models’ predictions \cite{Donoho2017}. By comparing the model's outputs to these outcomes, evaluators can compute a set of metrics (e.g., accuracy) to measure how well the system performs on a predetermined task. The evaluation process naturally mirrors training by repurposing the same data pipelines and directly measuring the model's performance on the task it was trained on \cite{kohavi1995cv}.

The task-specificity of these systems has three important implications for evaluations. First, evaluations of task-specific systems have \textbf{self-evident targets}. Because the model is trained with a clear objective, evaluators can measure how well the system performs its intended function.
Second, evaluations can be \textbf{properly scoped}. That is, evaluators can define the input space for each task-specific system and bound its functionality to specific use cases. This allows evaluators to assess model performance, knowing that the model will only operate within this defined input space and purpose.
Third, evaluations can be \textbf{reproducible}. When provided with the same inputs, outputs, and metrics, different evaluators can independently reproduce and verify the system's performance. These three properties allow policymakers to establish certain expectations around the validity and reliability of evaluations. Although there remain technical and institutional challenges in evaluations for task-specific systems, the expectations are that (i) evaluations should appropriately characterize the expected behavior of the systems relative to its defined objective (targeted), (ii) they should achieve sufficient coverage of the system's intended uses (well-scoped), and (iii) the evaluation results should be consistent and repeatable across different independent evaluators (reproducible). 

The absence of task specificity in GPAI systems means that these properties fall away. GPAI systems are not trained for specific tasks but rather with broader objectives to learn structural patterns in the data (e.g. predicting the next word given the previous words). As such, there is no clear target against which to evaluate their performance \cite{wallach2024evaluatinggenerativeaisystems}. Evaluators instead target general capabilities or properties, such as linguistic understanding \cite{wang2020supergluestickierbenchmarkgeneralpurpose}, reasoning \cite{chollet2019measureintelligence, cho2023dallevalprobingreasoningskills}, and biases \cite{parrish2022bbqhandbuiltbiasbenchmark, birhane2021multimodaldatasetsmisogynypornography, talat-etal-2022-reap}.
These capabilities or properties are characterized by the model's responses to a set of custom-built and annotated scenarios, typically in the form of multiple-choice or binary questions \cite{liang2023holisticevaluationlanguagemodels, srivastava2023imitationgamequantifyingextrapolating, hendrycks2021measuringmassivemultitasklanguage, wang2020supergluestickierbenchmarkgeneralpurpose}. 
A scenario is meant to instantiate a desired use case for GPAI models, and a collection of scenarios form benchmarks that guide and measure progress in AI development \cite{liang2023holisticevaluationlanguagemodels}.\looseness=-1

While capabilities-based evaluations are meant to produce evidence of likely performance on a broad range of possible tasks that would presumably draw on relevant capabilities, the relationship between capabilities and performance on specific, real-world tasks is often quite tenuous. As many have observed, the particular tasks used in capabilities-focused benchmarks don’t often resemble the tasks that GPAI models are used to perform in practice \cite{subramonian2023takestangonavigatingconceptualizations, raji2021ai, mcintosh2024inadequacieslargelanguagemodel, davis2023benchmarksautomatedcommonsensereasoning, bowman-dahl-2021-will, ethayarajh2021utilityeyeusercritique,mccoy2019rightwrongreasonsdiagnosing, ribeiro-etal-2020-beyond}. 

Even when tasks are well-specified, the real-world performance and impact of GPAI models are shaped by how those tasks are carried out via fine-tuning, prompting, and other forms of downstream adaptations \cite{raji2021ai, liao2023rethinking, bowman-dahl-2021-will, chi2025copilotarenaplatformcode}. These post-training modifications are often highly variable and unpredictable \cite{betley2025emergentmisalignmentnarrowfinetuning}, leading the same model to exhibit markedly different behavior even when used to perform what seem to be the same or similar tasks. Given how sensitive evaluation results can be to these downstream choices, it can be exceedingly difficult to reliably predict performance based on evaluations that only consider a limited set of possible choices.

Evaluation at the foundation model level falls prey to what Selbst et al. call the "framing trap"---the failure to account for the broader sociotechnical system in which AI operates \cite{selbst2019fairness}. 
With GPAI, the framing trap is largely unavoidable if one seeks to evaluate the foundation model. The sociotechnical context of the AI system varies with each use case, shaped by the distinct patterns of user interactions, institutional norms, and task-specific objectives \cite{blodgett-etal-2020-language}. As a result, even if it is feasible to develop robust, standardized evaluation methods for the foundation model itself, evaluating models in isolation from the context of its uses cannot adequately surface concrete harms or support meaningful mechanisms of accountability, which would be seen or apply only to the \emph{combination} of model and application context.

While past regulatory requirements around AI audits and impact assessments have hinged on the assumption that developers can meaningfully evaluate AI systems prior to deployment, GPAI challenges the reliability of current evaluation methods as regulatory tools. One example that highlights its limitations
is California’s S.B. 1047. The vetoed state bill would have required model developers ``to assess whether the covered model is reasonably capable of causing or materially enabling critical harm'' before making it publicly available \cite{california2024sb1047}. The problem, however, is that evaluating harm at the model level, without accounting for downstream use cases, fails to consider the role of user interactions, implementation decisions, and the organizational processes in shaping the real-world impacts of model use. As such, regulatory proposals hitched to pre-deployment evaluations of GPAI models may misdiagnose where and how harms actually materialize, leading to ineffective or misguided oversight. 
\subsection{General-purpose AI opens up new areas of legal concern that were not implicated by task-specific AI.}
\label{legal areas}

Recent developments make clear that GPAI poses legal challenges largely absent from task-specific AI. As such, the responses of existing legal regimes will necessarily differ, each requiring distinct domain-specific expertise.
Copyright offers the most prominent example to the general public. While task-specific AI can raise copyright concerns at the input stage---whether training on protected works constitutes infringement \cite{levendowski2018copyright, sobel2017fairuse}---this was not the primary legal concern with the technology. GPAI not only magnifies concerns about infringement in training, but also implicates the reproduction right due to its ability to output text, images, and videos, which task-specific AI did not \cite{lee2024talkinboutaigeneration, henderson2023foundation, lemley2023generative}. AI companies currently face 
major infringement lawsuits based on using protected content in model training and producing outputs that closely resemble copyrighted works \cite{grynbaum2023nyt, authorsguild2023openai, chabon2023openai, getty2023complaint}.

Beyond copyright concerns, GPAI has drawn scrutiny for its role in spreading defamation and misinformation. GPAI systems regularly fabricate plausible-sounding falsehoods, producing content with invented names, fake quotations, and non-existent sources \cite{openai2023}. Already, lawsuits have been filed over alleged libel created by GPAI programs \cite{walters_v_openai2024, battle_v_microsoft2024}. As GPAI systems are increasingly integrated into search engines and virtual assistants, the risk of reputational harm and public deception intensifies. In more extreme cases, GPAI systems have been used to commit fraud. Synthetic voice technologies have been used to clone speech patterns of real individuals, enabling impersonation through robocalls \cite{fcc2024aivoice, barrington2025voiceclones}. Hyperrealistic deepfake videos and AI-generated images have similarly been used to deceive viewers and distort perception, and pose particular difficulties for evidentiary processes in courts \cite{nightingale2022faces, bohacek2025aianchor}. The proliferation of synthetic content raises the question of whether and when AI developers should be held liable for content generated by its systems \cite{volokh2023libel, lee2024talkinboutaigeneration}.\looseness=-1

While task-specific systems have long played a role in intelligence analytics and surveillance, GPAI introduces new concerns in chemical, biological, radiological, and nuclear (CBRN) security \cite{ukgov2024aiseoulsummit, nist2024ai6001}. Traditionally, the development of CBRN weapons has required high levels of expertise, access to restricted materials, and coordination across domains. However, GPAI systems could lower informational barriers and expand the capabilities of malicious actors by synthesizing vast bodies of literature and proposing novel ideas for CBRN weapon development \cite{soice2023largelanguagemodelsdemocratize, mouton2024operational, gopal2023releasingweightsfuturelarge}.
While the actual realization of such threats still depends on real-world capabilities \cite{peppin2025realityaibiorisk}, the possibility that GPAI models could support CBRN threats has raised concerns about their dual-use potential, driving new discussions about risk mitigation, early warning systems, and access controls \cite{openai2024earlywarning, WhiteHouse2023EO, batalis2024anticipating, nasem2025age}.

The legal challenges posed by GPAI in new domains highlight the limitations of treating AI as a singular regulatory target. These issues fall under distinct legal regimes, each with its own doctrines, precedents, and norms that have developed over decades. For example, copyright law has long grappled with questions of authorship, originality, and fair use, and has developed specific doctrinal tests to differentiate between protected creative works and derivative copies \cite{usc102b}. CBRN security governance has been grounded in the precautionary principle, the practice of taking preventative action in the face of uncertainty. It has relied on nonproliferation regimes, material control measures, and international oversight to restrict the development and spread of high-risk threats \cite{un_bwc_history}. Each of these domains has also drawn on distinct communities of experts and stakeholders with different priorities, risk tolerances, and terminologies. The differences in how each community conceptualizes and addresses the challenges posed by GPAI mean that regulatory approaches that work well in one domain may be ineffective or counterproductive in another. As a result, effective governance for GPAI will require broader and more diverse domain knowledge than is required for task-specific AI alone. Attempts to consolidate these distinct legal challenges into a single, overarching AI policy framework risks not only oversimplifying the regulatory landscape but also overlooking the stakeholders and experts best positioned to guide policy decisions.

\subsection{General-purpose AI is built on a highly distributed value chain that complicates accountability.}
\label{sec:supply}

AI systems are developed and deployed through complex, interdependent networks of data providers, compute providers, model developers, hosting services, model adapters, and application developers \cite{ lee2024talkinboutaigeneration, khan2024code, bommasani2023ecosystemgraphssocialfootprint}.
Collectively, these actors comprise the "AI value chain," a sequence of modular processes that "contribute towards the production, deployment, use, and functionality of AI technologies" \cite{cobbe2023accountability}. This modularity allows for specialization and efficient division of labor \cite{porter1985competitive}, but it also disperses control and obscures lines of responsibility \cite{widder2022dislocated, brown2023accountability, cen2023aisupplychains}.

Task-specific AI systems typically exhibit a tightly integrated value chain, from data curation to feature selection to model development.
The system's single purpose and structure are understood at the time of creation, and the creation itself is often done collaboratively between developer and deployer \cite{desilva2022ai}. Thus, the universe of potential harms is either known, or at least foreseeable, to both the developer and deployer of the system.

Generative AI complicates this picture. As model development is no longer tied to specific tasks, providers focus on building large, incredibly complex, general-purpose models that are designed to be customized and reused. Then, in the minimally complex case, a deployer will fine-tune the system for their use case and then deploy it. There are several important implications to this change for legal concepts related to responsibility and accountability.

The first is that the conceptual distance between the model developer's work and the ultimate harm is much larger. With a task-specific model, developers will be aware of the final use, and will often work in concert with their customers; now, with a GPAI model, they are building general-purpose technologies, and many ultimate uses will be hard, if not impossible, to foresee \cite{widder2022dislocated}. Additionally, the complexity of these models renders specific outcomes likely unforeseeable, even where the general domain of use can be predicted in advance \cite{cobbe2023accountability}. Foreseeability, in turn, is a bedrock limitation to legal liability in most instances \cite{hart1985causation}. This translates to a general aversion to holding creators of general-purpose systems with substantial lawful uses liable for downstream harms \cite{sony1984}. Telephone creators are not held liable for crimes organized over the phone, nor are computer makers for harms researched and planned on laptops.\looseness=-1

The second is that there are more---and more specialized---roles in the overall value chain. This is not inherently difficult for a legal regime to handle. For example, the EU's AI Act lays out different sets of responsibilities for providers, deployers, importers, and distributors \cite{EUAIAct2024}. But new classes of intermediate actors---model adapters and integrators---are getting involved. Model adapters fine-tune or modify GPAI models for specific tasks or domains, while integrators incorporate models into applications or systems, shaping how users interact with and experience them. The fluidity of the different value chains and the disaggregation of responsibilities into different roles can make it difficult for a statute to stay current \cite{engler2022valuechain}.\looseness=-1

This fragmentation of roles is not inherently problematic for a legal regime. In product liability, for instance, all parties in the chain of commerce are held liable for an injury to a plaintiff, and the court lets them sort out who ultimately pays. In the EU AI Act, a deployer who makes substantial modifications or changes the purpose of a deployed system becomes a provider as a legal matter \cite[Article 25]{EUAIAct2024}. As long as \emph{someone} is liable in the end, then oversight can be accomplished. What's different here is that lines between actors in the GPAI value chain are often blurry, making the specific questions of how responsibility is divided more difficult and raising the cost of oversight \cite{yew2024break}. Sticking with products liability as an example, under the so-called ``component parts doctrine,'' a supplier of a component in a larger products is liable for the harm only if the component itself is defective, the integration causes the component to be defective, or the maker of the component substantially participates in the integration \cite{ali1998productliability}. This doctrine relies on a continued conceptual separation between the component and the larger product after integration. What's different now is that the downstream uses actually \textit{morph} the product at issue. Once another actor fine-tunes a model, adapts it to a new context, or integrates it into a larger application, the resulting system cannot be cleanly disaggregated back into its original components to determine who is responsible for which aspects of its behavior. This lack of separability means that courts will need to examine in detail how the system was modified, what changes those modifications introduced, and how those changes are connected to the harm at issue. Untangling this highly convoluted chain of modifications and actors will raise both the duration and expense of litigation. \looseness=-1

Third, the extra roles and distributed global nature of the AI value chain may pose jurisdictional difficulties in legal accountability. Models can be developed in one country and be ported across jurisdictions trivially. Yet the legality of their training or deployment may differ by jurisdiction. For example, if personal data of any EU residents was used to train GPAI models, the training likely violated the GDPR, so what should EU regulators make of someone who independently fine-tunes one of the existing models for a local purpose? Indeed, Yew et al. \cite{yew2025redteamingaipolicy} argue that companies are likely to exploit the geographical scoping in the AI Act to continue to provide services to customers in the EU while avoiding governance requirements in the Act. Current US-EU disputes over whether the Digital Services Act should apply to US-based LLM providers, and its implication for free speech protections within the United States are yet another example of the cross-border jurisdictional challenges that GPAI presents.
These challenges are amplified in the case of open-source or open-weight models, which are published and freely accessible anywhere. In contrast, these kinds of concerns are less common with task-specific models, as they are typically bespoke systems trained on proprietary data and developed with infrastructure confined to a single jurisdiction.

None of this means that upstream providers should be exempt from responsibility for downstream harms.
For one, they maintain access to core system components, and retain the ability to control the systems they release. Because there are only a small number of upstream providers, they have immense market power \cite{cma2023foundationmodels}. Further, GPAI has an unprecedented demand for compute and data. While earlier task-specific systems might be trained on thousands or millions of examples, GPAI models rely on billions or trillions of data points \cite{ainow2023landscape}. On average, training large-scale GPAI models requires roughly 100 times more compute than earlier systems \cite{sevilla2022compute}. As a result, upstream providers have become systematically important within the GPAI value chain, as downstream actors increasingly rely on a few providers for access to pretrained models, large-scale data, and advanced compute infrastructure---resources foundational to building and scaling GPAI systems \cite{cobbe2023accountability}. These intensified dependencies give upstream providers an outsized influence over who can build, scale, and commercialize GPAI, a dynamic that was far less pronounced in the context of task-specific AI.\looseness=-1
\section{Recommendations}
Whether emerging technology drives policy has been a longstanding question in technology law, including AI governance \cite{crootof2021structuring, calo2015robotics, jones2018does}. Drawing on past lessons from tech policymaking, we propose three recommendations for policymakers to consider when it comes to GPAI. First, regulators should identify regulatory targets based on the \textit{function} and \textit{consequences} of AI systems rather than technical implementation. Second, only in \textit{narrow circumstances} should the technical properties of AI systems be employed to determine the level of regulatory scrutiny. Third, regulators should examine the entire \emph{risk chain} of AI applications and strengthen structural constraints that could serve as more effective barriers to misuse.\looseness=-1

\subsection{Regulators should generally identify regulatory targets based on the applications and consequences of AI systems rather than technical implementation.}
\label{rec1}
 
The challenge of regulating AI exemplifies many of the core sources of contention and difficulties in tech policy. Central to this discourse is the distinction between tech-neutral and tech-specific regulatory approaches. Tech-neutral regulation governs activities and consequences, while tech-specific regulation governs specific technical implementations. 


The case for tech-neutral regulation has taken on new significance in the context of AI. In task-specific AI, model development is oriented to specific use cases: this task specificity facilitates compliance with tech-neutral regulation, as we argue in Section \ref{sec:generality}. However, for GPAI, this specificity is lost due to the separation between development and deployment. Instead, the general-purpose abilities of GPAI have led to calls for regulation at the level of the model itself, leading to shallow approaches based on model parameters and other technical proxies. 
\citet{bommasani2023tiers} has identified two types of proxies commonly used in AI policy: (1) resource expenditure (i.e., the amount of compute in FLOPs and the amount of data in modality-specific units), and (2) model properties (i.e. capabilities benchmarks and safety benchmarks).\looseness=-1

As the AI landscape evolves, however, rigid regulatory boundaries defined by technical proxies such as compute thresholds or model size quickly lose relevance.  While these metrics offer an initial assessment of the model's capabilities, they are limited. Technical proxies represent static snapshots that may neither reflect changes in model capabilities over time, nor behaviors that arise from user interactions in real-world scenarios.  In fact, we already see systems that achieve state-of-the-art capabilities with significantly reduced compute needs and dramatic performance gains from simple modifications in adaptation procedures \cite{deepseekai2025deepseekr1incentivizingreasoningcapability}, rendering such proxies ineffective as regulatory boundaries \cite{villasenor2025deepseek, zhang2025deepseek, chen2025deepseek}. By directing regulatory attention to the deployment context and intended use, regulators can better calibrate the level of regulatory scrutiny to the application-specific risks while accommodating ongoing technical progress.

AI regulation that relies on technical proxies risks not just becoming outdated but being both under- and over-inclusive \cite{crootof2021structuring}.\footnote{Considered more precisely, this is a point about legal rules and standards, not tech neutrality \cite[p.~406]{crootof2021structuring}. Where regulations are more rule-like, they can be over- and under-inclusive. We make the point here about tech-specificity because we think it is likely that tech-specificity leads to rules more often than standards. This is both because the tendency for tech-specific regulation is already motivated by a desire for easier adminsitrability and because technical standards and specifications are often stated in quantitative terms. Examples of tech-specific standard-based regulation may become more prevalent if governments, for example, mandate that LLMs have output filters that accomplish some goal like fairness or a more complicated set of goals in the vein of Anthropic's attempt at ``Constitutional AI'' \cite{bai2022constitutionalaiharmlessnessai}, but for now, we think the assumption holds.}
Regulation is under-inclusive when it leaves out systems that achieve similar concerning outcomes through alternative methods. For instance, relying on compute thresholds to identify frontier AI models may inadvertently exclude systems that cause harm through less resource-intensive methods (e.g., misinformation bots) \cite{newsom2024veto}. Regulation is over-inclusive when it captures more systems than intended. For example, the same compute thresholds could sweep in benign applications like climate modeling that happen to share high computational demands. The challenge in defining the regulatory scope is compounded by the broad range of contexts in which the same AI systems can be deployed to serve vastly different purposes \cite{harris2021ai}. The generality and adaptability of GPAI therefore underscore the need for regulatory frameworks that account for the specific contexts and outcomes of their use \cite{murdick2020aidefinitions}.\looseness=-1 

Tech-neutral regulation also helps address the challenge of regulatory arbitrage, where actors might switch to alternative technical approaches to circumvent tech-specific rules while achieving the same concerning outcomes \cite{crootof2021structuring, burk2016perverse}. This is especially relevant given the rapid pace of AI development, where new architectures and methodologies emerge frequently. Tying regulations to specific technologies creates incentives for companies to design their systems in ways that fall outside the regulatory scope \cite{yew2025redteamingaipolicy}. 

One example that highlights the limits of tech-specific regulation is the European Union's legislation on genetically modified organisms (GMO) \cite{eu2001directive}. The legislation targets crops that have been altered by recombinant DNA technology to address public concerns over potential health risks. In response, seed producers have turned away from recombinant DNA technology toward mutagenesis, using mutagenic chemicals or radiation treatments to induce random genetic changes in plants \cite{kaskey2013mutant}. While these methods are not considered recombinant DNA technology, they achieve the same outcome of producing genetically modified crops with commercially desirable traits. Ironically, mutagenic crops may pose comparable or even greater risks than GMO crops on food safety and the environment \cite{kaskey2013scariest, morris2007eu}. This case demonstrates how tech-specific regulation can incentivize the development of what Dan Burk calls ``perverse innovations'' that ``ingeniously dodge the intended outcome of regulation, while still formally adhering to the text of the regulation'' \cite{burk2016perverse}.\looseness=-1


\subsection{Identifying regulatory targets by technical specifications can still make sense as a last resort.}
\label{4.2}
In recent years, the AI regulatory landscape has seen a chorus of concerns over AI safety and security risks \cite{tobin2024aipioneers, lovely2024laws, tufekci2025dangerous, pause2023openletter}. These concerns have led to a surge in tech-specific proposals that aim to regulate AI systems, with model properties as technical proxies for their capacity to precipitate harm in society \cite{EUAIAct2024, house2024enforce, WhiteHouse2023EO, california2024sb1047}.\looseness=-1 

Despite its limitations, tech-specific regulation has garnered support in AI oversight due to its feasibility, political expediency, and capacity for preemptive intervention \cite{ohm2010technology, crootof2021structuring}. Compute governance, for instance, establishes concrete thresholds, the amount of computing resources used to train a model, which can be easily verified and applied across different actors and systems \cite{sastry2024computingpowergovernanceartificial}. This establishes objective, measurable intervention points in AI development, providing clarity as to whether and how regulation applies to an actor or artifact and increasing consistency in the application of the law. Tech-specific approaches also enable regulators to precisely target oversight toward certain types of technology while minimizing the burden on others. By promulgating separate rules for AI models with different computational requirements and specifications, regulators can establish clear standards for compliance and enforcement without considering the breadth and depth of AI applications, each with its own legal implications and risk profile. 
Finally, tech-specific regulation allows for intervention before potentially harmful systems are developed or deployed. Regulation through technology can be appropriate where the goal is to make something impossible rather than merely illegal \cite{lessig2006codev2, reidenberg1998lex}. This is particularly important for AI systems that could pose catastrophic risks or irreversible damage, where relying on ex post remedies may be inadequate \cite{bengio2023written}.\looseness=-1

However, as we highlighted in the prior section, tech-specific regulation is not without its limitations. Its application should be carefully constrained to avoid unintended consequences, such as overregulating benign use cases and creating regulatory loopholes. For this reason, we recommend that technical proxies should be employed for regulation only in narrow circumstances where three conditions are met: (1) tech-neutral regulation is demonstrably impractical in addressing the risk of concern, (2) the risk being addressed is severe, plausible, and attributable to the existence and configuration of the technology, and (3) the benefits from tech-specific regulation outweigh the potential harms. We now expand on each of these three conditions.

First, tech-specific regulation should only be used when tech-neutral regulation is impossible or impractical. A common critique of tech-neutral regulation is that it may be ineffective and even impossible to implement, by the 
time harms of a technology become apparent through widespread adoption \cite{gee2009late}. 
For instance, in nuclear safety regulation, the long-term environmental and health impacts from radioactive contamination and nuclear meltdowns are so severe and irreversible that a tech-neutral approach is inadequate \cite{nrc2011safetyculture}. Instead, nuclear safety regulation relies on highly specific rules tailored to the unique risks of nuclear technology, such as strict access to nuclear materials and safety protocols that govern nuclear facilities \cite{worldnuclear2021safeguards}. Similarly, GPAI presents a severe consequence that task-specific AI does not: risk of catastrophic harm \cite{hendrycks2023overviewcatastrophicairisks}. Catastrophic harm refers to events with potentially devastating consequences for human society, such as the misuse of AI to create bioweapons or the loss of control over advanced AI systems \cite{sunstein2007catastrophic}. Unlike more predictable risks, catastrophic harms are difficult to quantify because they often involve scenarios with little to no historical data, making it nearly impossible to assign a precise probability of occurrence \cite{farber2011uncertainty}. Some caution that the race among companies to advance AI heightens the risk that these models could pose unforeseen threats, calling for pauses in AI development \cite{pause2023openletter}. The concern that GPAI might be used for a broad range of dangerous and difficult-to-anticipate applications necessitates tech-specific regulation to preemptively mitigate these risks rather than reacting after the fact, when it may be too late to avert disaster. \looseness=-1

Second, tech-specific regulation should be reserved for cases where the potential harm is plausible and evidence-based, significant, and attributable to the specific technology or technological configuration. An example is AI-generated nonconsensual intimate imagery (NCII) based in the likeness of real people, otherwise known as deepfakes. AI-generated NCII is a current problem, with significant and well-documented harm \cite{mcglynn2017beyond, henry2020image, laird2024deeptrouble}. While NCII is starting to be addressed with criminalization, civil liability, and/or mandated takedowns \cite{takeitdown2024}, NCII may also be a candidate for tech-specific regulation because the harm is accomplished immediately, cheaply, and at scale, making it harder to detect and remove \cite{thiel2023generative, iwf2024aiupdate}. An example of such tech-specific regulation would be a requirement that image generators have filters that prevent the generation of NCII with features of real people \cite{nist2024ai6001}. This is tech-specific because it regulates through a configuration of the technology, rather than addressing the harm directly, like liability for its creation or takedown requirements do \cite{caSB926_2024, maH4241_2024}. So far, the technological limitations have not been totally successful---ability to evade them being one of the risks of tech-specific regulation---but something could be created that performs better, and successfully prevents the harm \cite{ding2025malicioustechnicalecosystemexposing}.\looseness=-1

Third, tech-specific regulation may be justified when the risks posed by a specific technology are significant enough that the benefits of regulation interventions outweigh the costs. One example of this could be facial recognition technology (FRT). Hartzog and Selinger have argued that FRT's mere existence is uniquely dangerous in a way that justifies banning the technology outright, due in part to the unique role of the face in our identity and social interactions \cite{hartzog2018facial}. In fact, some state and local governments have already enacted bans on its use \cite{sheard2022movement}. These restrictions curb the indiscriminate deployment of facial recognition technology by state actors, to prevent mass privacy violations, wrongful arrests, and disproportionate targeting of marginalized communities. While facial recognition technology may offer potential benefits in security and efficiency, ban advocates argue that these advantages are grossly overshadowed by the severe and evidenced risk it poses \cite{harwell2020uighur, hill2020facial}. GPAI may present analogous concerns in biosecurity.
It risks facilitating a bio-engineered pandemic by (1) lowering the barrier to entry for non-experts to develop biological weapons, and (2) discovering novel, more lethal pathogens that evade current safeguards. The former may reduce the cost and expertise required to create harmful biological agents, while the latter raises the ceiling of harm by increasing the lethality and transmissibility of pathogens. These two outcomes establish distinct threat models and introduce different trade-offs that regulators will need to evaluate and prioritize when weighing the benefits of AI advancements for bioresearch against the costs of potential misuse.

Regardless of how narrowly regulatory targets are defined, regulators must maintain the flexibility to update the basis for legal scrutiny \cite{bommasani2023tiers}. Even proponents of compute governance acknowledge that regulatory thresholds---such as a training compute threshold of $10^{26}$ operations---or list-based controls on technologies can quickly become outdated as technology evolves \cite{sastry2024computingpowergovernanceartificial}. This necessitates establishing mechanisms, from the outset, to periodically re-evaluate and revise policies. Importantly, this does not mean updating a single threshold, but rather ensuring frameworks evolve in response to new developments and evidence \cite{carp2018autonomous}. To provide this adaptability in AI governance, Bommasani et al. have called for forward-looking blueprints that map societal conditions (e.g., specific AI capabilities) to appropriate policy responses \cite{bommasani2024path}.\looseness=-1

\subsection{Regulators should focus on structural constraints and dependencies that prevent harm.} 

The current discourse on AI regulation has predominantly focused narrowly on the AI value chain. While this has yielded concrete technical governance proposals \cite{openai2024earlywarning, sastry2024computingpowergovernanceartificial, browne2024aisafety}, it neglects the systemic nature of AI risks and impacts, as its harm manifests through interactions across an even broader ecosystem. We propose an evaluation framework to assess regulatory approaches across domains, using biosecurity as a case study. The framework examines where existing regulations assume conditions that GPAI now disrupt, and pinpoints where new measures are needed to close new gaps.\looseness=-1

One way to conceptualize the interactions between actors, systems, and resources across an ecosystem is through a risk chain: a sequence of actions that traces how potential harms are conceived, developed, and realized across an ecosystem. In biosecurity, a risk chain might begin with an actor having an “irresponsible, misguided, or malicious” intention \cite{Sandberg2020}. This intention gives rise to a biological idea, which is then converted into biological data---such as a pathogen genome---and transformed into a live biological artifact. The artifact is subsequently cultured, tested, and ultimately dispersed into the target environment \cite{Sandberg2020}. In the following, we use bioweapons development as a case study to study how harm progresses or is prevented from one stage to another.

Scholars have long observed that the particular arrangements and affordances of digital technology create extra-legal structures and contours that shape conduct in practice, often in ways that regulation and laws adapt to and build on \cite{reidenberg1998lex, lessig2006codev2, Surden2007, Ohm2018}. These structural ``constraints'' function as ``regulators of behavior that prevent conduct through technological or physical barriers in the world'' \cite{Surden2007}. These constraints can be physical (e.g., access to infrastructure), economic (e.g., material costs), or epistemic (e.g., domain expertise).\footnote{We note that these categories of barriers are not fixed or mutually exclusive. In practice, they often overlap and reinforce one another. For instance, a lack of technical knowledge (epistemic) may require hiring experts (economic), which in turn may be necessary to operate specialized systems (physical). While analytically distinct, these barriers often operate together as part of a broader regulatory environment.} Much like law, these constraints can make certain activities prohibitively difficult or expensive to carry out at scale. They shape which harms can propagate along a risk chain and which are effectively contained by the ecosystem.\looseness=-1

Bioweapons development offer a clear example of how structural constraints operate within a risk chain. Even if an actor begins with a harmful biological idea, progressing through the chain requires access to biomaterials and specialized lab infrastructure \cite{peppin2025realityaibiorisk, Batalis2023}. This requirement imposes physical constraints for obtaining and handling hazardous biomaterials and highlights risk chain dependencies, as the movement and use of biomaterials necessarily involve interactions with institutions that produce and distribute these agents \cite{ASPR_Biosafety}. These dependencies make certain transitions in the risk chain, such as transforming a digital genome into a cultured pathogen, contingent on navigating the physical, economic, and epistemic barriers. At the same time, these points offer opportunities for policy interventions, including screening and monitoring the distribution of biomaterials \cite{Kane2024, ASPR_SynDNA}, requiring licenses for researchers working with high-risk biological agents \cite{Jonsson2013}, and enforcing strict security protocols in laboratory facilities \cite{NRC2009}. \looseness=-1

We propose a three-step domain-independent framework to help identify and evaluate how AI interacts with the risk chain.
\textbf{First, regulators should identify the key dependencies and constraints at each stage of the risk chain.} These could include both deliberately engineered dependencies that make certain actions infeasible without authorization, as well as by-product constraints of the technological and physical environment raising the difficulty, cost, or time required to carry out harmful activities \cite{Surden2007}. In biosecurity, multiple constraints and dependencies impede the progression of the risk chain, making it more difficult to carry out biological attacks. To being, there are epistemic barriers due to a lack of biological knowledge and scientific training \cite{NRC2004, OuagrhamGormley2014}. Next, actors need biomaterials, specialized equipement, state-of-the-art methods and deep domain expertise \cite{Leitenberg2012}. And the final stages of culturing, testing, and deploying biological agents are time- and resource-intensive \cite{NRC2004, EC2005,NTI2023}

\textbf{Second, regulators should assess the extent to which existing regulations depend on structural constraints to remain effective.} When these constraints shift or weaken due to technological changes, the regulatory regimes built around them may lose their force. In the case of biosecurity, regulatory efforts to deter the creation of bioweapons have traditionally relied on physical constraints, such as controlling access to hazardous biomaterials and enforcing safety protocols within high-containment laboratory facilities \cite{un_bwc_history, ASPR_Biosafety, NSTC2022}. These efforts rest on the assumption that malicious actors cannot operationalize biological threats without overcoming significant physical barriers.

\textbf{Third, regulators should evaluate how advancing AI capabilities could undermine or bypass the structural constraints that existing regimes rely on.} Structural constraints lose their regulatory force when actors can leverage AI to circumvent the friction and barriers they impose at different stages of the risk chain. This could occur in three key ways. 
First, AI could collapse stages of the risk chain by streamlining multiple-step processes to a single, more efficient operation. In the biorisk chain, 
early stages of literature review, hypothesis formulation and experiment design can be bypassed entirely by LLMs fine-tuned on biological data \cite{NelsonRose2023, Batalis2023, Rose2024b, Allen2023}. 
Second, AI could discover new ways to formulate bioweapons using readily accessible materials, rather than known inputs that are subject to access controls \cite{Helena2023,sandbrink2023artificialintelligencebiologicalmisuse, soice2023largelanguagemodelsdemocratize, Batalis2023, luckey2025aiweapons},
thereby compromising biosecurity regimes that rely on physical containment. 
Third, AI could reduce friction by accelerating workflows, automating processes, and lowering the time and cost that would otherwise slow down or discourage certain activities \cite{boiko2023emergentautonomousscientificresearch, inagaki2023llmsgenerateroboticscripts,Rose2024b, soice2023largelanguagemodelsdemocratize}. 
These three mechanisms demonstrate how AI can reshape the structure and pace of the risk chain, potentially undermining constraints that once served as effective regulatory barriers.\looseness=-1

The extent to which these constraints remain effective depends on the regulatory objectives and assumptions underlying the structure of the risk chain. For instance, even if GPAI lowers the barrier to entry to biodevelopment---by expanding access to biological information and reducing the time and cost needed for pathogen development---the physical constraints likely remain intact \cite{Jefferson2014, Grushkin2013}. Alternatively, if regulation focused on epistemic constraints, such as restricting access to sensitive methodological knowledge by redacting details in published biology manuscripts \cite{Enserink2011}, then the regulatory emphasis would shift to limiting knowledge diffusion. Regulators should determine which constraints are most critical to preserving the function of existing legal regimes and assess how AI may undercut their regulatory force.
\looseness=-1 
\section{Conclusion}
Our goal in this work is to illustrate when the differences between general-purpose and task-specific AI are merely matters of degrees---where existing regulatory approaches remain applicable---and when they are differences in kind, demanding new regulatory instruments and forms of intervention. We identify four distinct areas of meaningful difference and make three recommendations for policymakers based on these differences. We believe that these recommendations can help policymakers avoid many of the fault lines in current AI policy debates and lead to more effective policymaking for GPAI.\looseness=-1
\newpage

\bibliographystyle{ACM-Reference-Format}
\bibliography{references}

\end{document}